\documentclass{jpsj2}

\title{%
The Triple Higgs Boson Self-Coupling at Future Linear $e^+e^-$
Colliders Energies: ILC and CLIC }

\author{%
A. Guti\'errez-Rodr\'iguez $^1$ \thanks{E-mail address:
alexgu@planck.reduaz.mx}, M. A. Hern\'andez-Ru\'iz $^2$, O. A.
Sampayo $^3$,\\
A. Chubykalo $^1$ and A. Espinoza-Garrido $^1$}

\inst{%
$^1$Facultad de F\'{\i}sica, Universidad Aut\'onoma de Zacatecas\\
         Apartado Postal C-580, 98060 Zacatecas, Zacatecas M\'exico.\\
$^2$Instituto de F\'{\i}sica, Universidad
        Aut\'onoma de San Luis Potos\'{\i}\\
        78000 San Luis Potos\'{\i}, SLP, M\'exico.\\
$^3$Departamento de F\'{\i}sica, Universidad Nacional del Mar del Plata\\
              Funes 3350, (7600) Mar del Plata, Argentina. }

\recdate{\today}

\abst{%
We analyzed the triple Higgs boson self-coupling at future
$e^{+}e^{-}$ colliders energies, with the reactions
$e^{+}e^{-}\rightarrow b \bar b HH, t \bar t HH$. We evaluate the
total cross-sections for both $b\bar bHH$ and $t\bar tHH$, and
calculate the total number of events considering the complete set
of Feynman diagrams at tree-level. We vary the triple coupling
$\kappa\lambda_{3H}$ within the range $\kappa=-1$ and +2. The
numerical computation is done for the energies expected to be
available at a possible Future Linear $e^{+}e^{-}$ Collider with a
center-of-mass energy $800, 1000, 1500$ $GeV$ and a luminosity
1000 $fb^{-1}$. Our analysis is also extended to a center-of-mass
energy $3$ $TeV$ and luminosities of 1000 $fb^{-1}$ and 5000
$fb^{-1}$. We found that for the process $e^{+}e^{-}\rightarrow b
\bar b HH$, the complete calculation differs only by $3\%$ from
the approximate calculation $e^{+}e^{-}\rightarrow ZHH(Z\to b\bar
b)$, while for the process $e^{+}e^{-}\rightarrow t \bar tHH$, the
expected number of events, considering the decay products of both
$t$ and $H$, is not enough to obtain an accurate determination of
the triple Higgs boson self-coupling.}

\kword{%
Total cross-sections; standard model Higgs boson. }

\begin{document}
\maketitle

\section{Introduction}

The Higgs boson ~\cite{Higgs,Higgs1,Higgs2} plays an important
role in the Standard Model (SM) ~\cite{SM,SM1,SM2} because it is
responsible for generating the masses of all elementary particles
(leptons, quarks, and gauge bosons). However, the Higgs-boson
sector is the least tested in the SM, in particular the Higgs
boson self-interaction.

The search for Higgs bosons is one of the principal missions of
present and future high-energy colliders. The observation of this
particle is of major importance for the present understanding of
fundamental particle interactions. Indeed, in order to accommodate
the well established electromagnetic and weak interaction
phenomena, the existence of at least one isodoublet scalar field
to generate fermion and weak gauge bosons masses is required.
Despite previous success in explaining the present data, the SM
cannot be completely tested before this particle has been
experimentally observed and its fundamental properties studied.

The triple and quartic Higgs boson couplings
~\cite{Boudjema,Ilyin,Djouadi,Djouadi1,Oslan,Boudjema1,Djouadi2,Djouadi3}
$\lambda_{3H}$ and $\lambda_{4H}$ are defined through the
potential:

\begin{equation}
V(H)=\frac{M^2_{H}}{2}H^2+\frac{M^2_{H}}{2v}H^3+\frac{M^2_{H}}{8v^2}H^4,
\end{equation}

\noindent where the triple and quartic couplings of the Higgs
field $H$ are given by

\begin{eqnarray}
\lambda_{3H}&=&\frac{3M^2_H}{M^2_Z}\lambda_0,\\
\lambda_{4H}&=&\frac{3M^2_H}{M^4_Z}\lambda^2_0.
\end{eqnarray}

To obtain these expressions we assumed the normalization employed
in Ref. ~\cite{Djouadi,Djouadi1,Oslan,Boudjema1,Djouadi2,Djouadi3
}, where $\lambda_0=M^2_Z/v$.

In the SM, we obtain $M_H= \sqrt{2\lambda}v$ as the simple
relationship between the Higgs boson mass $M_{H}$ and the
self-coupling $\lambda$, where $v=246$ $GeV$ is the vacuum
expectation value of the Higgs boson. The triple vertex of the
Higgs field $H$ is given by Eq. (2) and a measurement of
$\lambda_{3H}$ in the SM can determine $M_H$. An accurate test of
this relationship may reveal the extended nature of the Higgs
sector. The measurement of the triple Higgs boson coupling is one
of the most important goals of Higgs physics in a future $e^+e^-$
linear collider experiment. This would provide the first direct
information on the Higgs potential responsible for electroweak
symmetry breaking.

The triple Higgs boson self-coupling can be measured directly in
pair-production of Higgs particles at hadron and high-energy
$e^+e^-$ linear colliders. Several mechanisms that are sensitive
to $\lambda_{3H}$ can be exploited for this task. Higgs pairs can
be produced through double Higgs-strahlung of $W$ or $Z$ bosons
~\cite{Djouadi,Djouadi1,Oslan,Boudjema1,Djouadi2,Djouadi3,
Gounaris,Barger,Kamoshita,Miller,Miller1,Yasui,Battaglia}, $WW$ or
$ZZ$ fusion ~\cite{Ilyin,Barger1,Dobrovolskaya,Dicus,Abbasabadi};
moreover, through gluon-gluon fusion in $pp$ collisions
~\cite{Glover,Plehn,Plehn1,Dawson} and high-energy $\gamma\gamma$
fusion ~\cite{Ilyin,Jikia} at photon colliders. The two main
processes at $e^+e^-$ colliders are double Higgs-strahlung and
$WW$ fusion:

\begin{eqnarray}
\mbox{double Higgs-strahlung}&:& e^+e^- \to ZHH,  \nonumber \\
\mbox{$WW$ double-Higgs fusion}&:& e^+e^- \to \bar\nu_e \nu_e HH.
\end{eqnarray}

The $ZZ$ fusion process of Higgs boson pairs is suppressed by an
order of magnitude because the electron-Z coupling is small. The
more suitable reaction in $e^+e^-$ colliders to measure the triple
couplings in the range of the theoretically preferred Higgs mass
${\cal O}(100\hspace{1mm} $GeV$)$ is the double Higgs-strahlung
process $e^+e^- \to ZHH$. Operating at center-of-mass energy
$\sqrt{s}$ from $500$ $GeV$ up to about 1 $TeV$, the International
Linear Collider (ILC) ~\cite{Abe,Abe1} can measure the $HHZ$
production cross-section (about 0.1-0.2 $fb$) if the Higgs boson
mass is $M_H=120$ $GeV$ ~\cite{Castanier}. When the center-of-mass
energy $\sqrt{s}$ exceeds 1 $TeV$, the $e^+e^- \to \bar\nu_e \nu_e
HH$ mode becomes sizeable and it is possible to measure the triple
Higgs self-coupling $\lambda_{3H}$ by using this process.
Therefore, in the first stage of the ILC $(\sqrt{s}< 1
\hspace{1mm} $TeV$)$, $e^+e^- \to ZHH$ is the most promising
channel to measure the triple Higgs self-coupling $\lambda_{3H}$.
In this process, the final state of two Higgs may be generated by
the interchange of a virtual Higgs in such a way that this process
is sensitive to the triple coupling $HHH$ in the Higgs potential.
It is necessary to include four-body processes with heavy fermions
$f$, $e^{+}e^{-}\rightarrow f\bar f HH$, in which the SM Higgs
boson is radiated by a $b(\bar b)$ quark at future $e^{+}e^{-}$
colliders
~\cite{NLC,NLC1,JLC,A.Gutierrez,A.Gutierrez1,A.Gutierrez2,A.Gutierrez3,A.Gutierrez4
} with a c.m. energy in the range of 800 to 1500 $GeV$, as in the
case of the ILC ~\cite{Abe,Abe1} and Compact Linear Collider
(CLIC) ~\cite{CLIC} machines, in order to know its impact on the
three-body channel and also to search for new relations that may
have a clear signature of the Higgs boson production.

\begin{figure}[t]
\begin{center}
\includegraphics[width=12cm]{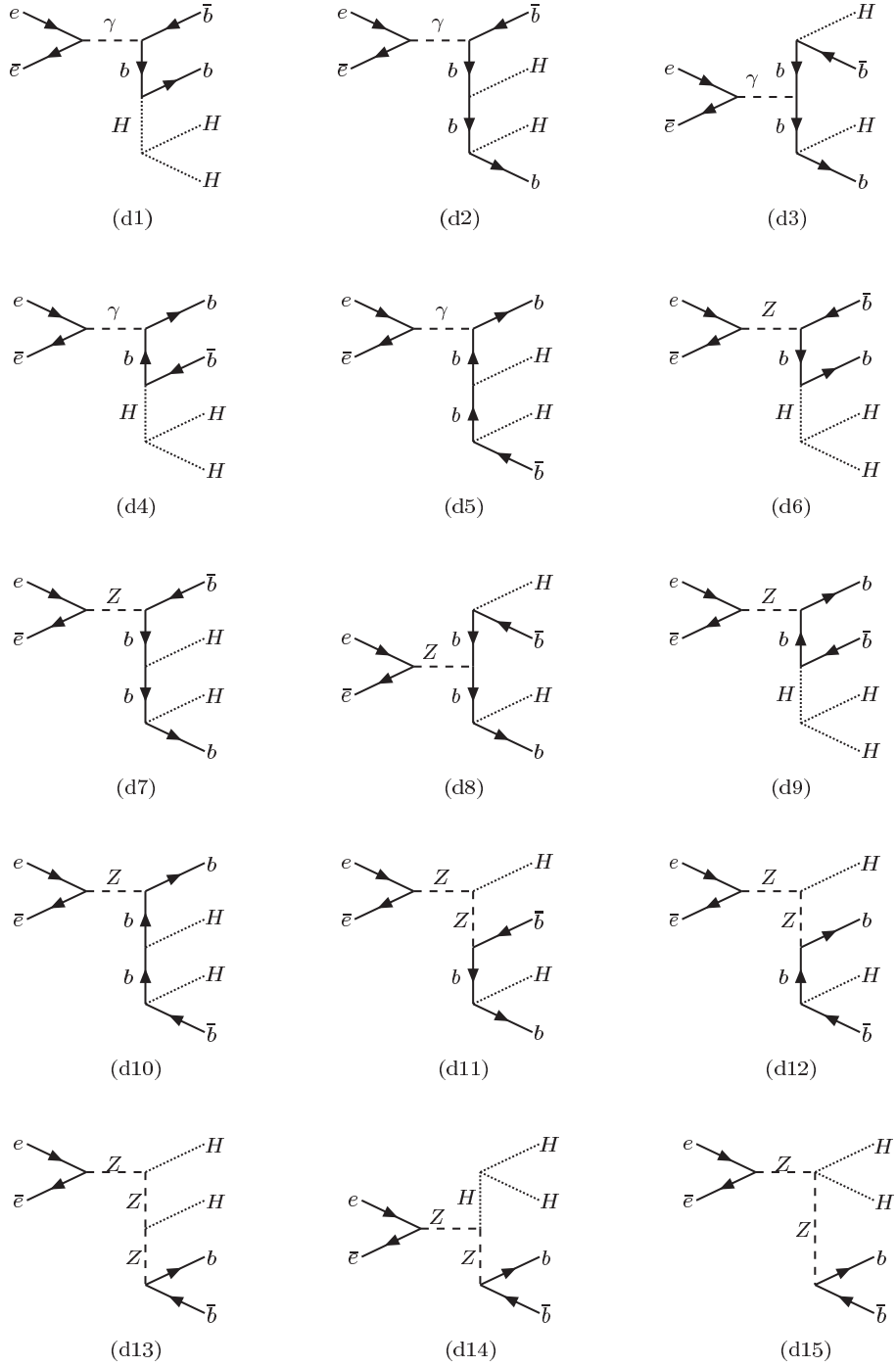}
\end{center}
\caption{ Feynman diagrams at tree-level for $e^{+}e^{-}
\rightarrow b\bar b HH$.} \label{f1}
\end{figure}

\begin{figure}[t]
\begin{center}
\includegraphics[width=12cm]{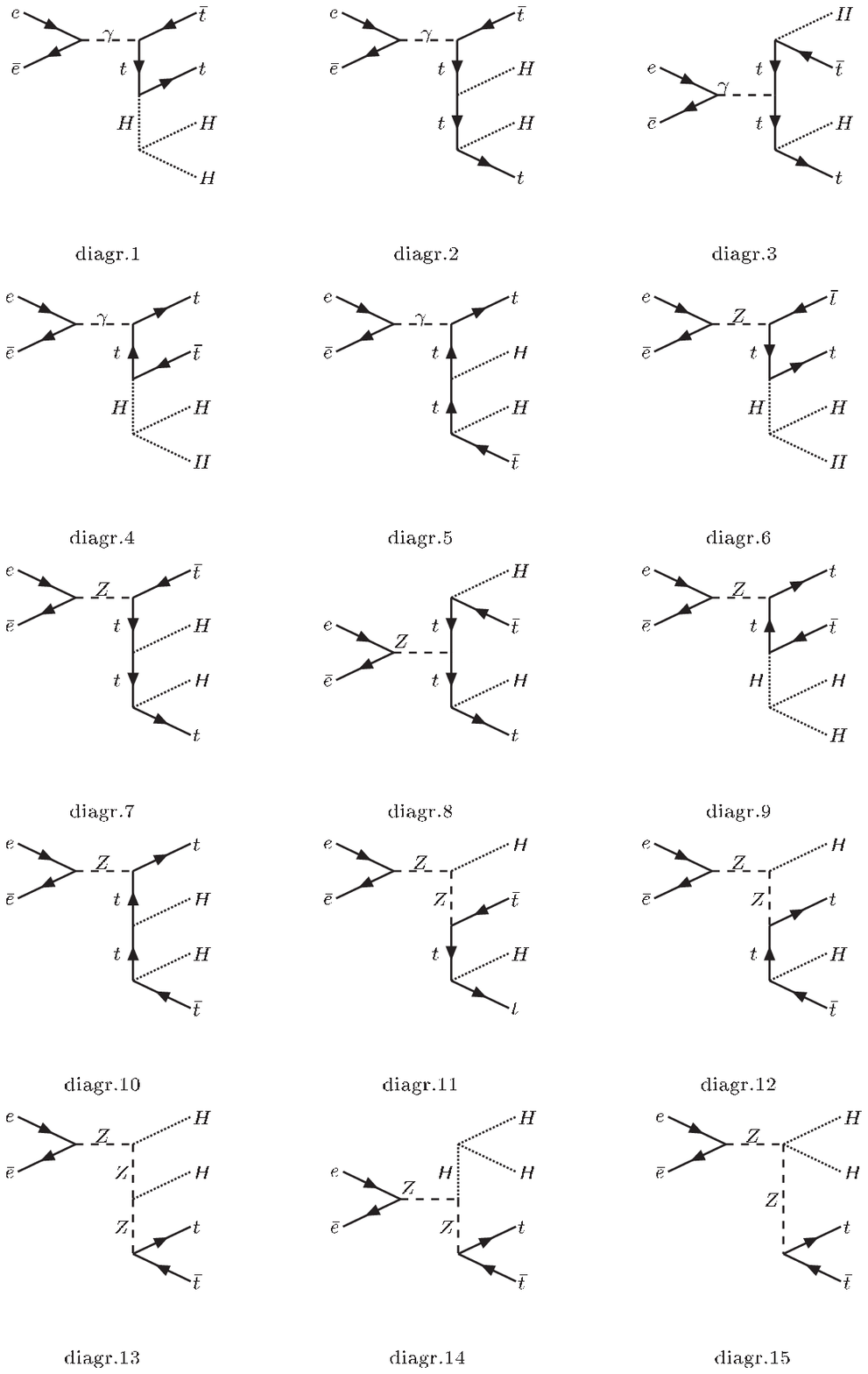}
\end{center}
\caption{ Feynman diagrams at tree-level for $e^{+}e^{-}
\rightarrow t\bar t HH$.} \label{f2}
\end{figure}

The Higgs coupling with top quarks, the largest coupling in the
SM, is directly accessible in the process where the Higgs boson is
radiated off top quarks, $e^{+}e^{-}\rightarrow t\bar t HH$. This
process depends on the Higgs boson triple self-coupling, which
could lead us to obtain the first non-trivial information on the
Higgs potential. We are interested in finding regions that could
allow the observation of the $b\bar bHH$ and $t\bar tHH$ processes
at future linear $e^{+}e^{-}$ colliders energies: ILC and CLIC. In
the process $e^{+}e^{-}\rightarrow b \bar b HH$, the set of
figures shown for the $b\bar bHH$ final state includes the $ZHH$
process with $Z\to b\bar b$. We found that the results for the
complete calculation $e^{+}e^{-}\rightarrow b \bar b HH$ and for
the approximate $e^{+}e^{-}\rightarrow ZHH$ with an on-shell $Z$
decay to $b\bar b$, differ only at the $3\%$ level in the examined
kinematic range. We consider the complete set of Feynman diagrams
at tree-level ( Figs.~\ref{f1} and ~\ref{f2}) and use the CALCHEP
~\cite{Pukhov} packages to evaluate the amplitudes and
cross-section of the processes $e^{+}e^{-}\rightarrow b \bar b HH$
and $e^{+}e^{-}\rightarrow t \bar t HH$.

This paper is organized as follows: In Sec. 2, we study the triple
Higgs boson self-coupling through the processes
$e^{+}e^{-}\rightarrow b \bar b HH$ and $e^{+}e^{-}\rightarrow t
\bar t HH$ at future linear $e^{+}e^{-}$ colliders energies and,
finally, we summarize our results in Sec. 3.\\

\section{Cross-Section of the Higgs Boson Pairs Production with Triple
Self-Coupling}
 In this section we present numerical results for
$e^{+}e^{-}\rightarrow b \bar b HH$ and $e^{+}e^{-}\rightarrow t
\bar t HH$ with double Higgs boson production. We carry out the
calculations using the Standard Model framework at future linear
$e^{+}e^{-}$ colliders energies. We use the CALCHEP ~\cite{Pukhov}
packages for calculations of the matrix elements and
cross-sections. These packages provide automatic computation of
the cross-sections and distributions in the SM as well as their
extensions at tree-level. We consider the high energy stage of a
possible future linear $e^{+}e^{-}$ collider with $\sqrt{s}=800,
1000, 1500$ $GeV$ and the designed luminosity 1000 $fb^{-1}$.

\subsection{Triple Higgs Boson Self-Coupling Via $e^{+}e^{-}\rightarrow b \bar b HH, t \bar t HH$}

\begin{figure}[t]
\begin{center}
\includegraphics[width=9cm]{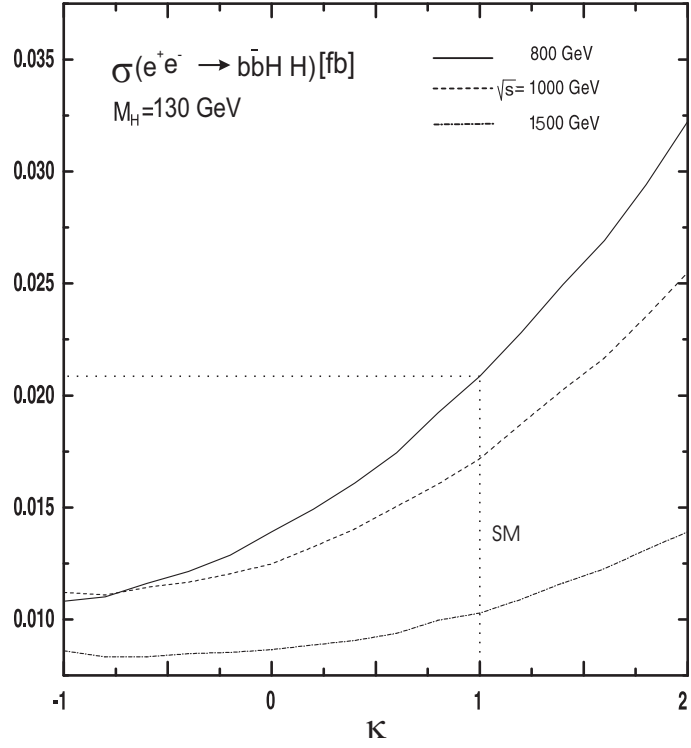}
\end{center}
\caption{ Variation of the cross-section $\sigma(b\bar b HH)$ with
the modified triple coupling $\kappa\lambda_{3H}$ at a collider
energy of $\sqrt{s}= 800, 1000, 1500$ $GeV$ and $M_H= 130$.}
\label{f3}
\end{figure}

\begin{figure}[t]
\begin{center}
\includegraphics[width=10cm]{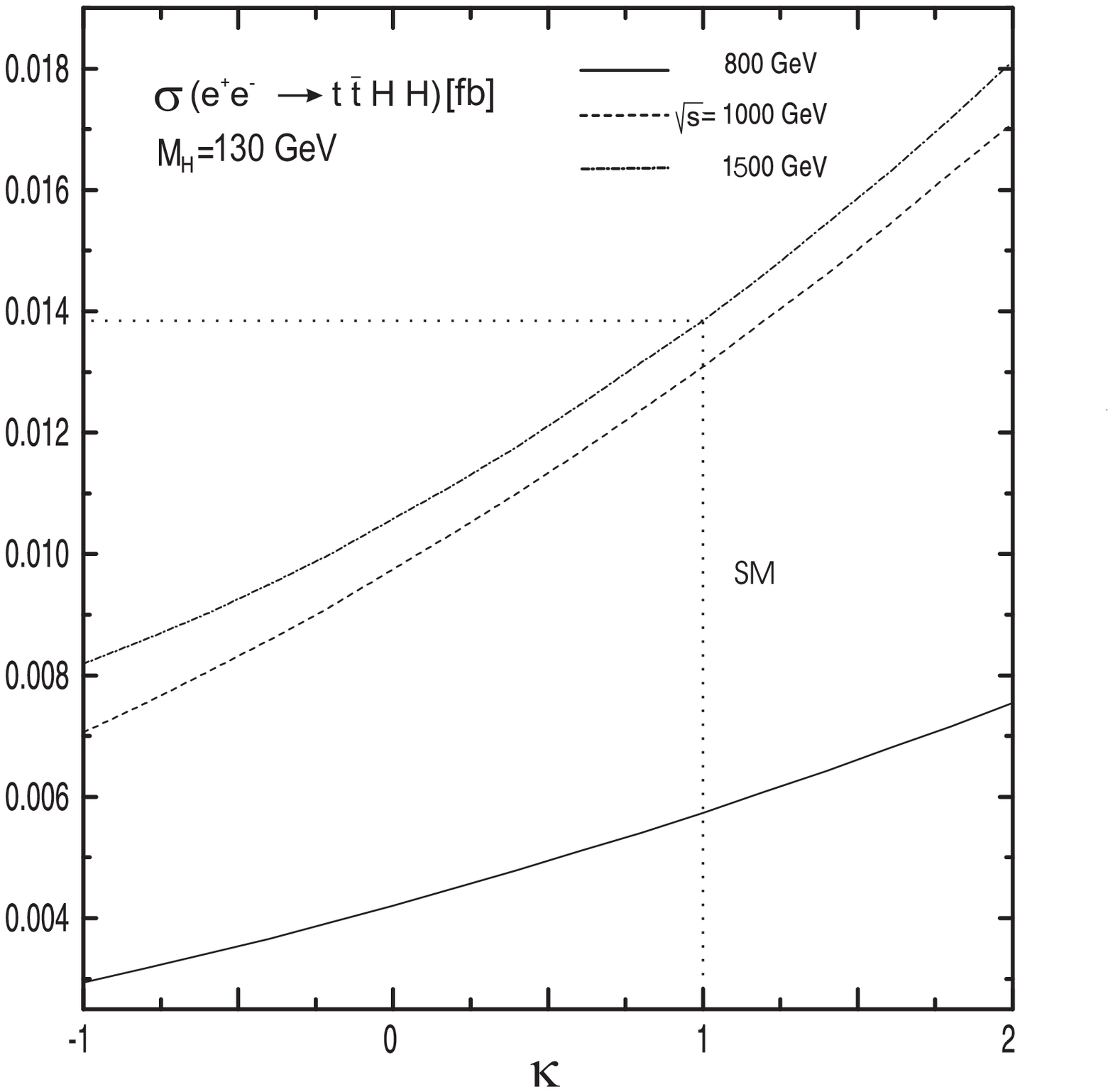}
\end{center}
\caption{ The same as in Fig. 3, but for the process $e^{+}e^{-}
\rightarrow t\bar t HH$.} \label{f4}
\end{figure}

\begin{figure}[t]
\begin{center}
\includegraphics[width=9cm]{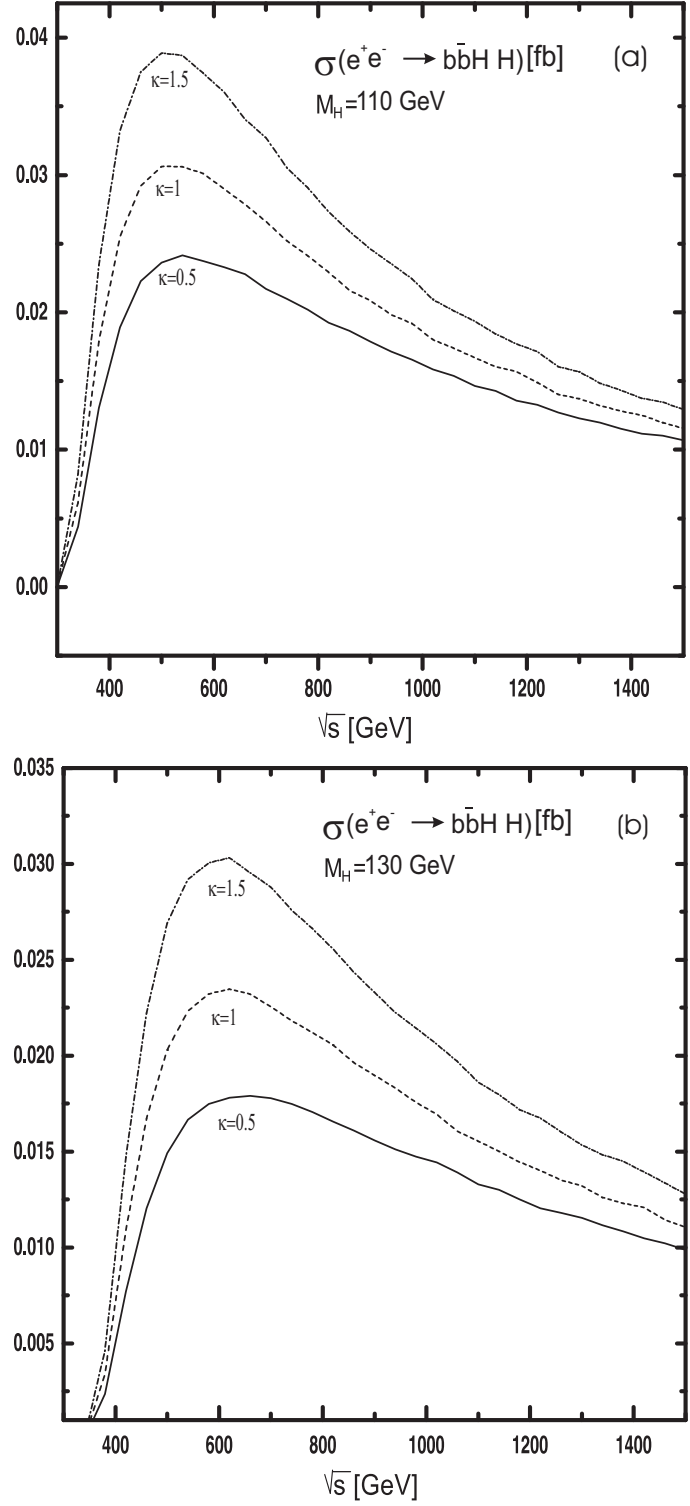}
\end{center}
\caption{ The dependence of the cross-section on center-of-mass
energy $\sqrt{s}$ for two fixed Higgs masses $M_H= 110, 130$
$GeV$. The variation of the cross-section for modified triple
couplings $\kappa \lambda_{3H}$ is indicated by the solid and
dot-dashed lines.} \label{f5}
\end{figure}

\begin{figure}[t]
\begin{center}
\includegraphics[width=10cm]{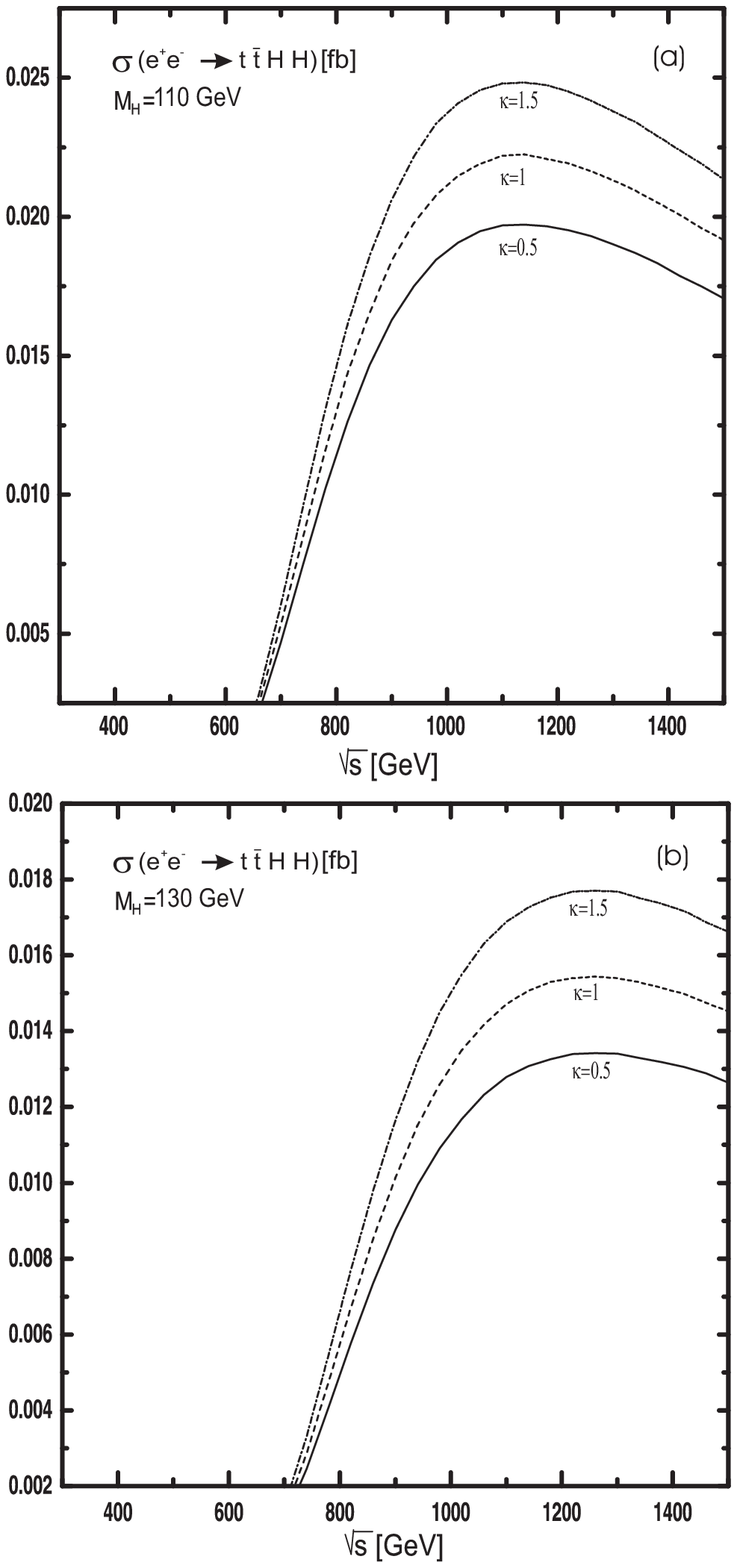}
\end{center}
\caption{ The same as in Fig. 5, but for the process $e^{+}e^{-}
\rightarrow t\bar t HH$.} \label{f6}
\end{figure}

In order to illustrate our results for the sensitivity to the
$HHH$ triple Higgs boson self-coupling, we show the $\kappa$
dependence of the total cross-section for $e^{+}e^{-}\rightarrow b
\bar b HH$ in Fig.~\ref{f3} and for $e^{+}e^{-}\rightarrow t \bar
t HH$ in Fig.~\ref{f4}. We consider one representative value of
the Higgs boson mass, $M_H=130$ $GeV$, with a center-of-mass
energy of $\sqrt{s}=800, 1000, 1500$ $GeV$, varying the triple
coupling $\kappa \lambda_{3H}$ within the range $\kappa=-1$ and
$+2$. In both cases, the cross-section is sensitive to the value
of the triple coupling. The sensitivity to $\lambda_{3H}$
increases with the collider energy, reaching a maximum at
$\sqrt{s}\sim 600$ $GeV$ for the $b \bar b HH$ channel and at
$\sqrt{s}\sim 1200$ $GeV$ for the $t \bar t HH$ channel
(Figs.~\ref{f5} and ~\ref{f6}). As an indicator of the order of
magnitude, in Tables~\ref{t1} -~\ref{t3} we present the number of
events of Higgs bosons expected for several Higgs boson masses,
center-of-mass energy and $\kappa$ values and for $1000 fb^{-1}$
luminosity (of course, we have multiplied by the corresponding
Branching Ratios to obtain the observable number of events). In
particular, if we consider the $H\rightarrow b \bar b$ decay for
$M_H < 130$ $GeV$, there is some possibility to detect the
$e^{+}e^{-}\rightarrow b \bar b HH$ process. In this region, the
number of events is small but sufficient to detect
$e^{+}e^{-}\rightarrow b \bar b HH\rightarrow b \bar b b \bar b b
\bar b$, in which the $BR(H\rightarrow b \bar b)\sim 0.6$ and the
background for 6 b-jet are small.

For the $e^{+}e^{-}\rightarrow t \bar t HH$ process, the
center-of-mass energy $1000$ $GeV$ and $M_H < 130$ $GeV$ is the
most favorable, but the Branching Ratios for the four decay modes
make this process very small.

For the center-of-mass energy $\sqrt{s}$ from 800 $GeV$ up to
about 1 $TeV$, the production of $b \bar b HH$ and $t \bar t HH$
in the intermediate mass range of the $H$ mass is significant and
all the final states can be identified without large momentum
loss. When the c.m. energy $\sqrt{s}$ exceeds 1 $TeV$, the
cross-section decreases and therefore in the first stage of a
future International Linear Collider $(\sqrt{s}\leq 1 $TeV$)$,
$e^{+}e^{-}\rightarrow b \bar b HH$ and $e^{+}e^{-}\rightarrow t
\bar t HH$ are important channels to measure the triple Higgs
boson self-coupling.

\begin{figure}[t]
\begin{center}
\includegraphics[width=12cm]{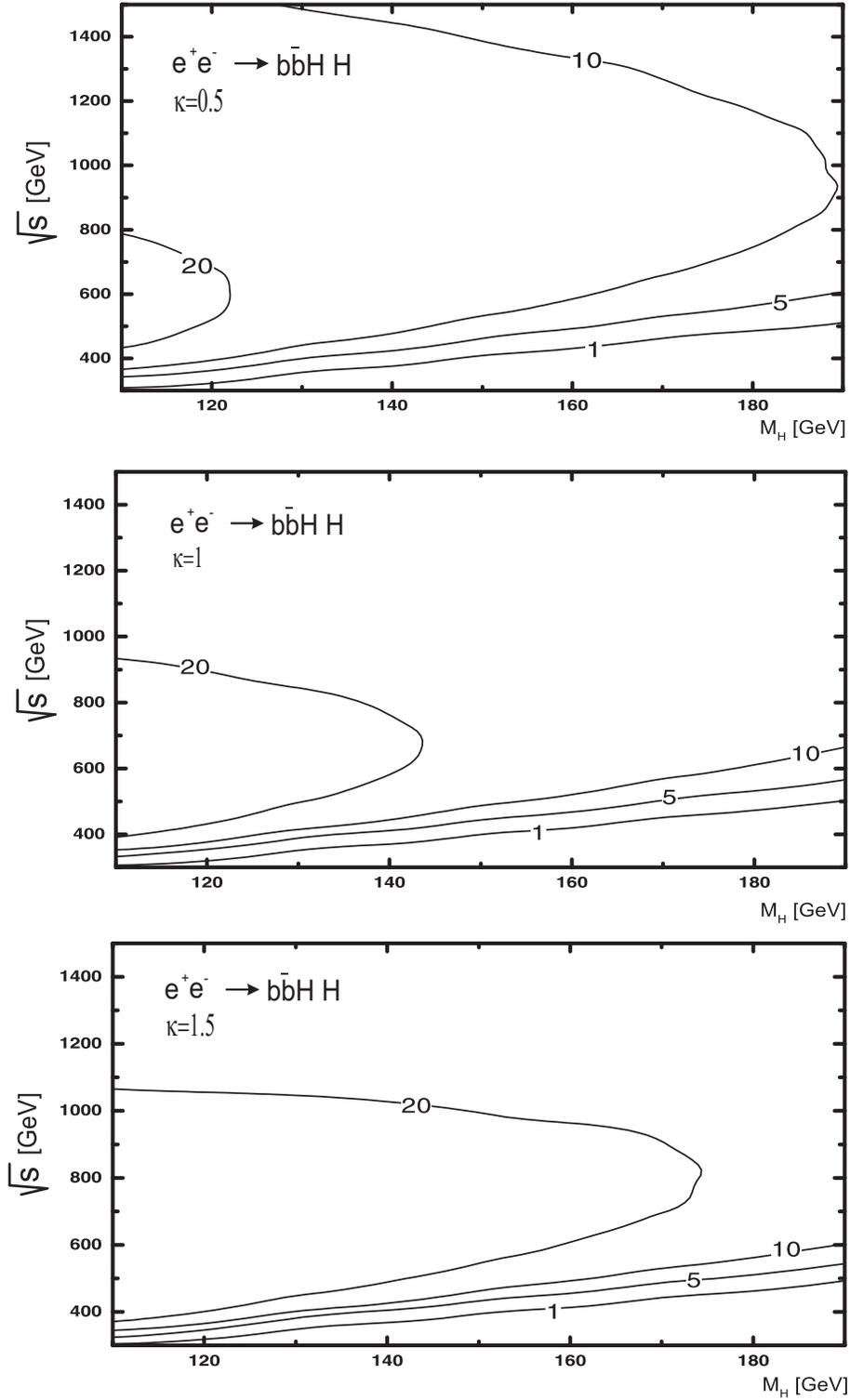}
\end{center}
\caption{ Contour plot for the number of events of the process
$e^{+}e^{-}\rightarrow b\bar b HH $ as a function of $M_H$ and
$\sqrt{s}$. The variation of the number of events for modified
triple couplings $\kappa\lambda_{3H}$ is indicated for $\kappa=
0.5, 1(S.M.), 1.5$.} \label{f7}
\end{figure}

\begin{figure}[t]
\begin{center}
\includegraphics[width=12cm]{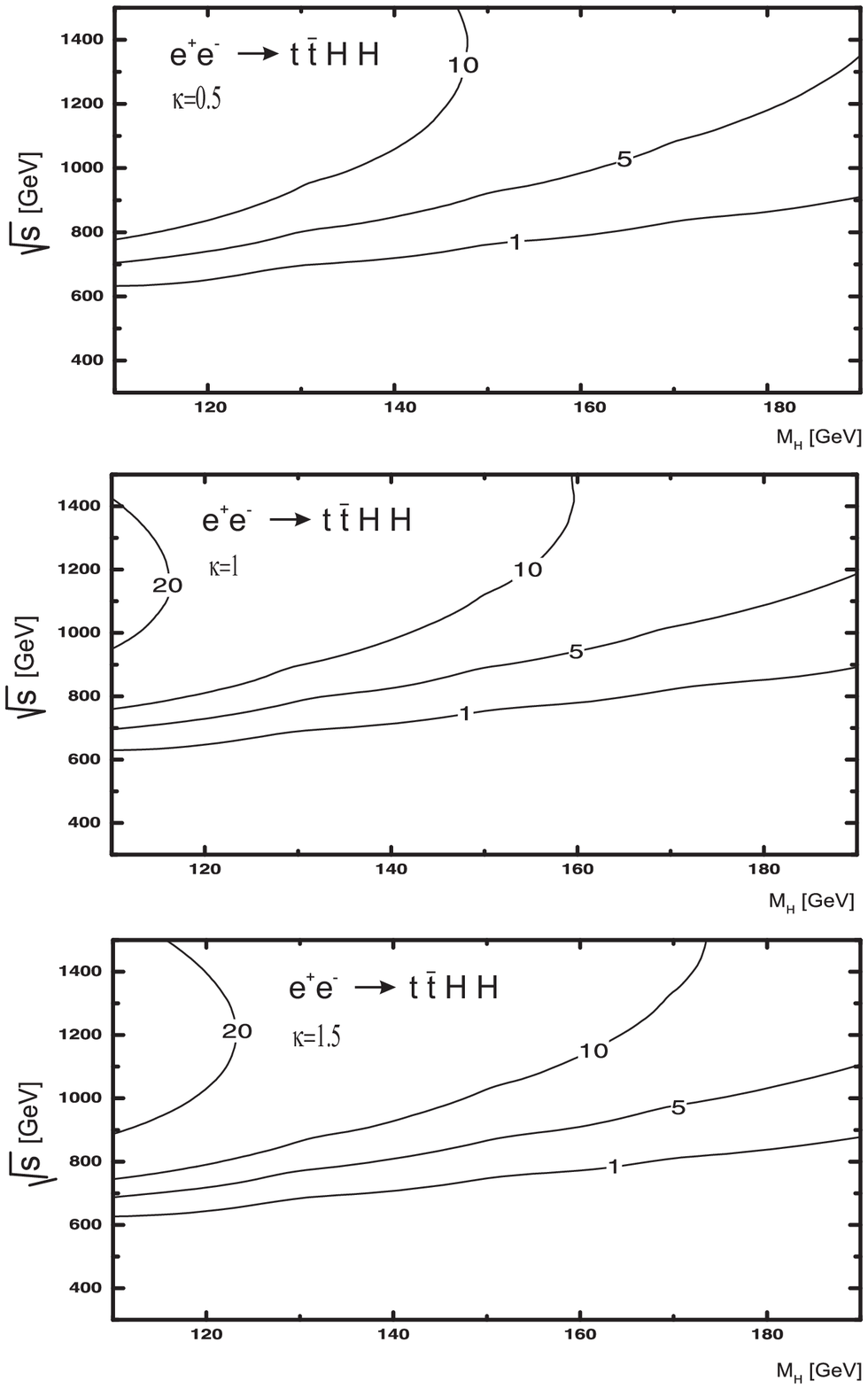}
\end{center}
\caption{ The same as in Fig. 7, but for the process $e^{+}e^{-}
\rightarrow t\bar t HH$.} \label{f8}
\end{figure}

Finally, we include a contour plot for the number of events of the
studied processes as a function of $M_H$ and $\sqrt{s}$ with
$\kappa= 0.5, 1(S.M.), 1.5$ in Figs.~\ref{f7} and ~\ref{f8}. These
contours are obtained from Tables~\ref{t1} -~\ref{t3}.

\begin{table}[t]
\caption{Total production of Higgs boson pairs in the SM for
${\cal L}=1000$ $fb^{-1}$ and $\kappa=0.5$.}
\label{t1}
\begin{center}
\begin{tabular}{c|c|c|c}
\hline
Total Production of Higgs Boson Pairs & \multicolumn{3}{c}{$e^{+}e^{-}\rightarrow b \bar b HH(t \bar t HH) \hspace{8mm} \kappa=0.5$}\\
\hline \hline
\cline{2-4} & $\sqrt{s}= $ & $\sqrt{s}= $ & $\sqrt{s}= $  \\
$M_H(GeV)$ & 800 $GeV$ & 1000 $GeV$ & 1500 $GeV$ \\
\hline \hline
 110 & 20 (11) & 16 (18) & 10 (17) \\
 130 & 17 (5)  & 14 (11) &  9 (13) \\
 150 & 14 (2)  & 12 (6)  &  9 (9)  \\
 170 & 11 (-)  & 11 (4)  &  8 (7)  \\
 190 &  9 (-)  & 10 (2)  &  8 (5)  \\
\hline
\end{tabular}
\end{center}
\end{table}

\begin{table}[t]
\caption{Total production of Higgs boson pairs in the SM for
${\cal L}=1000$ $fb^{-1}$ and $\kappa=1(SM)$.}
\label{t2}
\begin{center}
\begin{tabular}{c|c|c|c}
\hline
Total Production of Higgs Boson Pairs & \multicolumn{3}{c}{$e^{+}e^{-}\rightarrow b \bar b HH (t \bar t HH) \hspace{8mm} \kappa=1(SM)$}\\
\hline \hline
\cline{2-4} & $\sqrt{s}= $ & $\sqrt{s}= $ & $\sqrt{s}= $  \\
$M_H(GeV)$ & 800 $GeV$ & 1000 $GeV$ & 1500 $GeV$ \\
\hline \hline
 110 & 23 (13) & 18 (21) & 12 (19) \\
 130 & 21 (5)  & 17 (13) & 11 (14) \\
 150 & 18 (2)  & 16 (8)  & 10 (11) \\
 170 & 15 (-)  & 14 (-)  & 10 (8)  \\
 190 & 13 (-)  & 13 (-)  & 10 (6)  \\
\hline
\end{tabular}
\end{center}
\end{table}

\begin{table}[t]
\caption{Total production of Higgs boson pairs in the SM for
${\cal L}=1000$ $fb^{-1}$ and $\kappa=1.5$.}
\label{t3}
\begin{center}
\begin{tabular}{c|c|c|c}
\hline
Total Production of Higgs Boson Pairs & \multicolumn{3}{c}{$e^{+}e^{-}\rightarrow b \bar b HH (t \bar t HH) \hspace{8mm} \kappa=1.5$}\\
\hline \hline
\cline{2-4} & $\sqrt{s}= $ & $\sqrt{s}= $ & $\sqrt{s}= $  \\
$M_H(GeV)$ & 800 $GeV$ & 1000 $GeV$ & 1500 $GeV$ \\
\hline \hline
 110 & 28 (15) & 21 (24) & 13 (20) \\
 130 & 26 (6)  & 21 (15) & 13 (16) \\
 150 & 23 (3)  & 20 (9)  & 13 (13) \\
 170 & 20 (-)  & 18 (5)  & 13 (10) \\
 190 & 17 (-)  & 17 (3)  & 13 (8)  \\
\hline
\end{tabular}
\end{center}
\end{table}

\subsection{Triple Higgs Boson Self-Coupling Via $e^{+}e^{-}\rightarrow b \bar b HH, t \bar t HH$ at CLIC Energies}

\begin{figure}[t]
\begin{center}
\includegraphics[width=8.5cm]{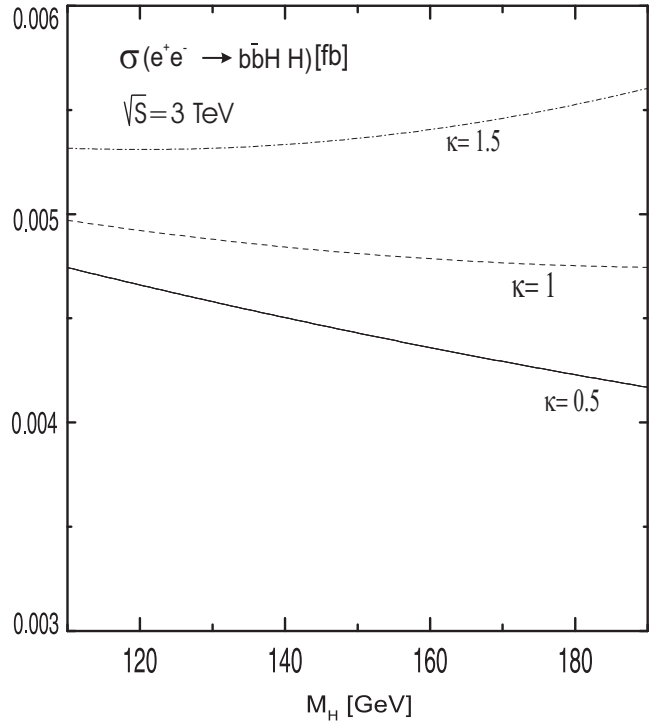}
\end{center}
\caption{ The cross-section for the double Higgs-strahlung via
$e^{+}e^{-}\rightarrow b \bar b HH$, at a c.m. energy of
$\sqrt{s}=3$ $TeV$ as a function of $M_H$ with $\kappa= 0.5,
1(S.M.),1.5$. The effects of a variation of the triple coupling by
$50\%$ from its SM value are shown.} \label{f9}
\end{figure}

\begin{figure}[t]
\begin{center}
\includegraphics[width=8.5cm]{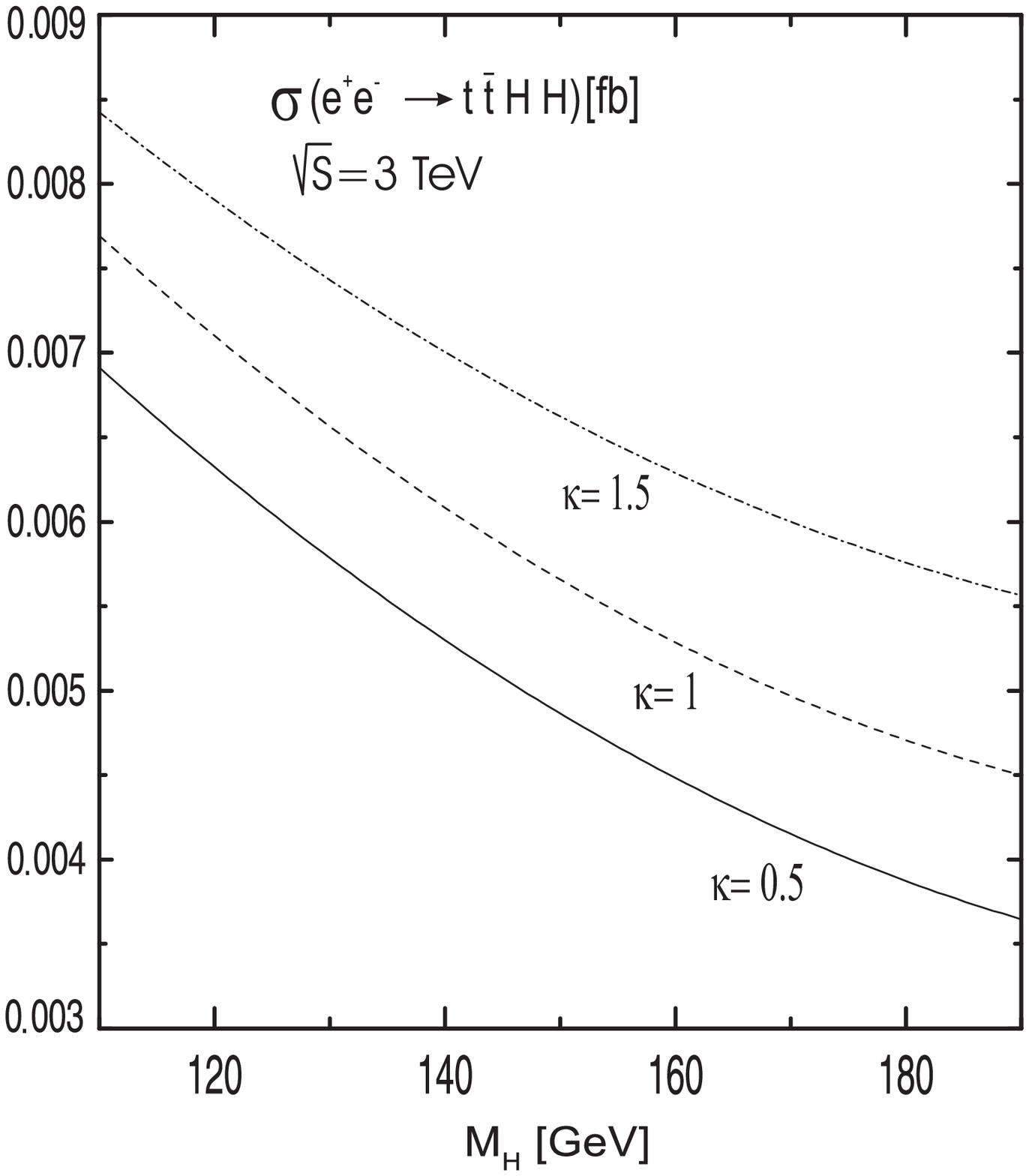}
\end{center}
\caption{The same as in Fig. 9, but for the process $e^{+}e^{-}
\rightarrow t\bar t HH$.} \label{f10}
\end{figure}

In this subsection we analyze the triple Higgs self-coupling
$\lambda_{3H}$ via the processes $e^{+}e^{-}\rightarrow b \bar b
HH, t \bar t HH$ for energies expected at the CLIC \cite{CLIC}.
Figs.~\ref{f9} and ~\ref{f10} show the total cross-section for the
double Higgs-strahlung in $e^{+}e^{-}$ collisions,
$e^{+}e^{-}\rightarrow b \bar b HH, t \bar t HH$ as a function of
$M_H$ for the c.m. energy of $\sqrt{s}=3$ $TeV$ and $\kappa= 0.5,
1(S.M.),1.5$. The effects of a variation of the triple coupling by
$50\%$ from its SM value are shown in these figures. The
production cross-section is of the order of a fraction of a
femtobarn ($0.005 fb$ for $b \bar b HH$ and $0.008 fb$ for $t\bar
t HH$) when it is not overly suppressed by phase-space and it is
mediated by $s$ channel gauge boson exchange. From these figures,
we observe that the total cross-sections of both processes are too
small because their order of magnitude is smaller than that for
the case of $\sqrt{s}=800, 1600$ $GeV$, as indicated in Ref.
\cite{A.Gutierrez,A.Gutierrez1,A.Gutierrez2,A.Gutierrez3,
A.Gutierrez4}.

\begin{figure}[t]
\begin{center}
\includegraphics[width=9cm]{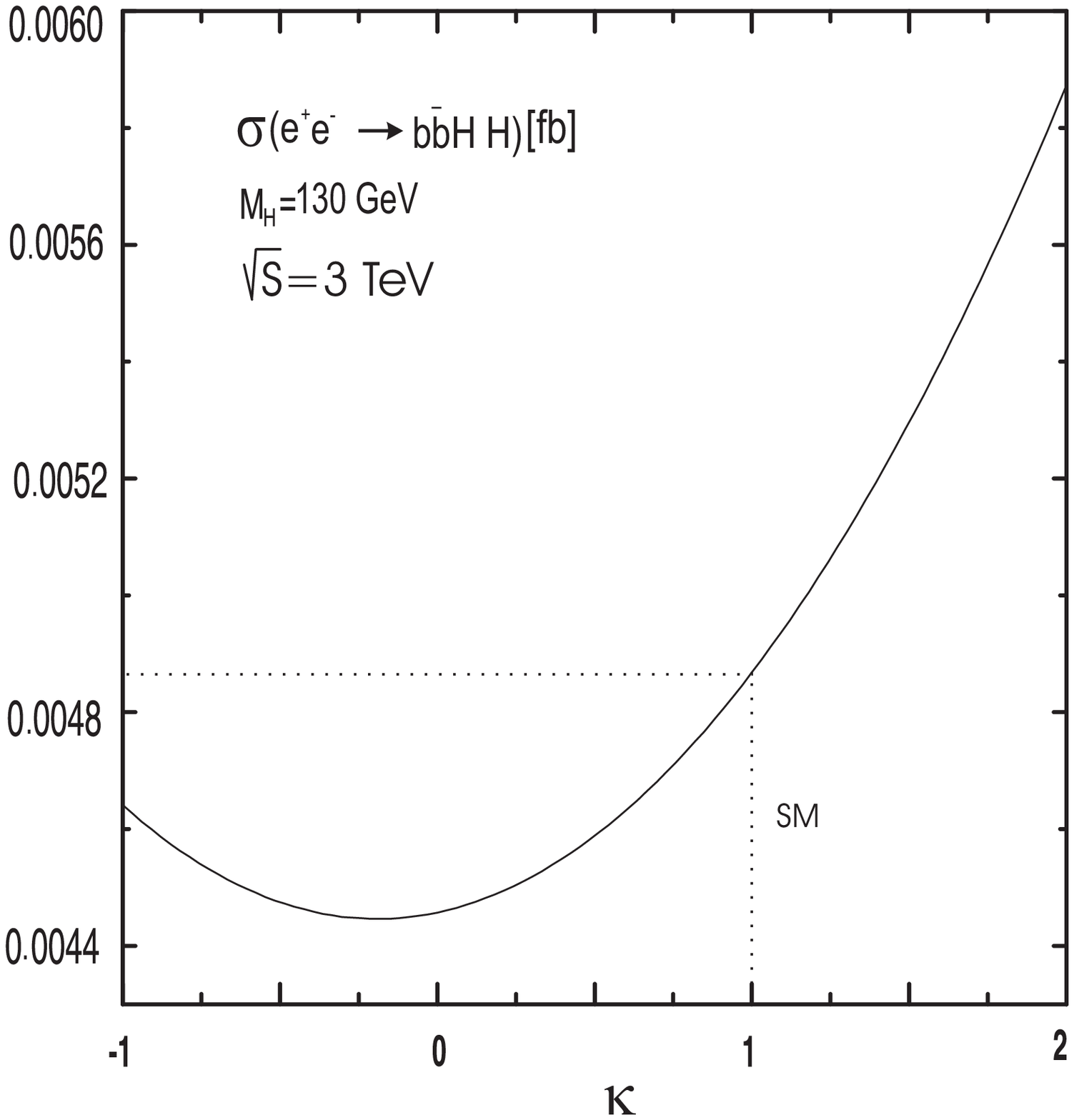}
\end{center}
\caption{Variation of the cross-section $\sigma(b\bar b HH)$ with
the modified triple coupling $\kappa\lambda_{3H}$ at a collider
energy of $\sqrt{s}=3$ $TeV$ and $M_H= 130$.} \label{f11}
\end{figure}

\begin{figure}[t]
\begin{center}
\includegraphics[width=9cm]{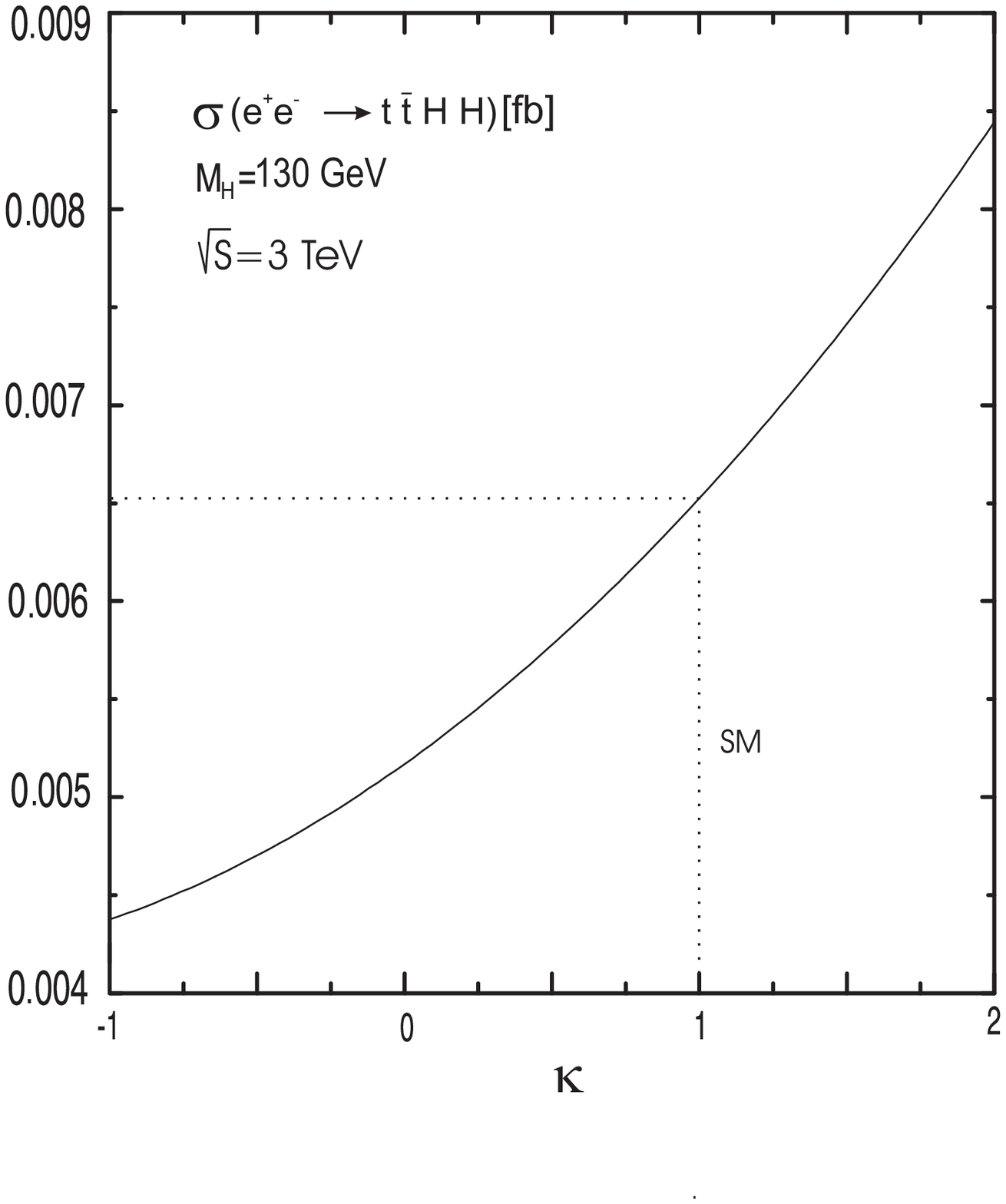}
\end{center}
\caption{ The same as in Fig. 11, but for the process $e^{+}e^{-}
\rightarrow t\bar t HH$.} \label{f12}
\end{figure}

As in subsection 2.1, we show the $\kappa$ dependence of the total
cross-section for $e^{+}e^{-}\rightarrow b \bar b HH, t \bar t HH$
in Figs.~\ref{f11} and ~\ref{f12}. We consider one representative
value of the Higgs boson mass, $M_H=130$ $GeV$, and center-of-mass
energy $\sqrt{s}=3$ $TeV$, varying the triple coupling $\kappa
\lambda_{3H}$ within the range $\kappa=-1$ and $+2$. In both
cases, the production cross-sections are also too small because
their order of magnitude is smaller than that for the case of
$\sqrt{s}=800, 1500$ $GeV$ and $M_H=130$ $GeV$, as is illustrated
in Figs.~\ref{f3} and ~\ref{f4} of subsection 2.1.

Finally, in  Tables~\ref{t4} and ~\ref{t5} we present the Higgs
boson number of events for several Higgs boson masses, $\kappa$
values, luminosities of $1000 fb^{-1}$ and $5000 fb^{-1}$ and
center-of-mass energy $\sqrt{s}=3$ $TeV$ (of course, we have
multiplied by the corresponding Branching Ratios to obtain the
observable number of events). It is clear from Figs.~\ref{f9}
-~\ref{f12} and Table~\ref{t4} that it would be difficult to
obtain a clear signal of the processes $e^{+}e^{-}\rightarrow b
\bar b HH, t \bar t HH$ at energies of a future linear collider
such as CLIC, after having considered the background, except for
$\sqrt{s}= 3$ $TeV$ and very high luminosity (${\cal L}= 5000$
$fb^{-1}$) as is shown in Table~\ref{t5}. However, for the
center-of-mass energy of CLIC, the $WW$ double Higgs fusion
process \cite{Ilyin,Boudjema,Castanier}, which increases with
rising $\sqrt{s}$, can be exploited by larger energies and
luminosities, and would be the preferred mechanism to measure the
triple Higgs self-coupling $\lambda_{3H}$.

\begin{table}[t]
\caption{ Total production of Higgs boson pairs in the SM for
$\sqrt{s}=3$ $TeV$ and ${\cal L}=1000$ $fb^{-1}$.}
\label{t4}
\begin{center}
\begin{tabular}{c|c|c|c}
\hline
Total Production of Higgs Boson Pairs & \multicolumn{3}{c}{$e^{+}e^{-}\rightarrow b \bar b HH (t \bar t HH), \hspace{8mm} \sqrt{s}=3$ $TeV$}\\

\hline \hline
\cline{2-4} $M_H(GeV)$ & $\kappa=0.5 $ & $\kappa=1 (SM) $ & $\kappa=1.5 $  \\
\hline \hline
 110 & 5 (7) & 5 (8) & 5 (8) \\
 130 & 5 (6) & 5 (7) & 5 (7) \\
 150 & 4 (5) & 5 (6) & 5 (7) \\
 170 & 4 (4) & 5 (5) & 5 (6) \\
 190 & 4 (4) & 5 (4) & 6 (6)  \\
\hline
\end{tabular}
\end{center}
\end{table}

\begin{table}[t]
\caption{Total production of Higgs boson pairs in the SM for
$\sqrt{s}=3$ $TeV$ and ${\cal L}=5000$ $fb^{-1}$.}
\label{t5}
\begin{center}
\begin{tabular}{c|c|c|c}
\hline
Total Production of Higgs Boson Pairs & \multicolumn{3}{c}{$e^{+}e^{-}\rightarrow b \bar b HH (t \bar t HH), \hspace{8mm} \sqrt{s}=3$ $TeV$}\\

\hline \hline
\cline{2-4} $M_H(GeV)$ & $\kappa=0.5 $ & $\kappa=1 (SM) $ & $\kappa=1.5 $  \\
\hline \hline
 110 & 24 (35) & 25 (39) & 27 (42) \\
 130 & 23 (29) & 24 (33) & 26 (37) \\
 150 & 22 (26) & 24 (28) & 27 (33) \\
 170 & 21 (21) & 24 (25) & 27 (30) \\
 190 & 21 (18) & 24 (22) & 28 (28)  \\
\hline
\end{tabular}
\end{center}
\end{table}

\section{Conclusions}

$e^+e^-$ linear colliders represent a possible opportunity for
triple Higgs boson self-coupling analysis. Therefore, we have
analyzed the triple Higgs boson self-coupling at future $e^+e^-$
collider energies with the reactions $e^{+}e^{-}\rightarrow b \bar
b HH$ and $e^{+}e^{-}\rightarrow t \bar t HH$. The ILC has access
to the triple Higgs boson self-coupling through the double Higgs
production processes $e^{+}e^{-}\rightarrow ZHH$ and
$e^{+}e^{-}\rightarrow \nu \bar \nu HH$
\cite{Boudjema,Ilyin,Djouadi,Djouadi1,Oslan,Boudjema1,Djouadi2,Djouadi3,
Kamoshita,Miller,Miller1,Yasui,Battaglia,Castanier}. Although the
cross section for $e^{+}e^{-}\rightarrow ZHH$ with intermediate
Higgs boson mass is only about 0.1-0.2 $fb$ for $\sqrt{s} < 1$
$TeV$, the measurement of the Higgs self-coupling $\lambda_{3H}$
at $e^{+}e^{-}$ colliders can be significantly improved. For
example, in Ref. \cite{Castanier}, C. Castanier et al. concluded
that a precision of about $10\%$ on the total cross-section for
$e^{+}e^{-}\rightarrow ZHH$ can be achieved, leading to a relative
error on $\lambda_{3H}$ of $18\%$ with the help of high integrate
luminosity ${\cal L}=2ab^{-1}$ after performing the detailed
simulations of signal and background process at the TESLA
\cite{TESLA}. Other simulations
\cite{Kamoshita,Miller,Miller1,Yasui} demonstrate that the Higgs
self-coupling $\lambda_{3H}$ can be extracted more accurately by
using some discriminating kinematic variables, namely the
invariant mass of the $HH$ system and the extraction of the Higgs
self-coupling $\lambda_{3H}$ can be further improved to an
accuracy of $8\%$ and even better in multi-$TeV$ $e^{+}e^{-}$
collisions \cite{Battaglia}. Therefore, to determine the triple
Higgs boson self-coupling $\lambda_{3H}$ via the process
$e^{+}e^{-}\rightarrow b \bar b HH$ and due to the cross-section
difference of $3\%$ between $ZHH (Z\to b\bar b)$ and $b\bar bHH$,
the conclusions for the precision in the determination of
$\lambda_{3H}$ are not significantly modified. That is to say, we
expect that the results for the complete computation of the
process $e^{+}e^{-}\rightarrow b \bar b HH$ should not alter the
conclusions of the previous computation
\cite{Kamoshita,Miller,Miller1,Yasui,Castanier} and thus the same
background analysis for $e^{+}e^{-}\rightarrow ZHH (Z\to b\bar b)$
remains valid for $e^{+}e^{-}\rightarrow b \bar b HH$. Examination
of variables sensitive to the triple Higgs boson vertex and the
availability of high luminosity will allow testing of the Higgs
potential structure at future linear $e^+e^-$ colliders (in the
case of the Minimal Supersymmetric extension of the Standard Model
(MSSM) with large $\tan\beta$, the $b\bar b HH$ channel may be
significantly enhanced). On the other hand, for the $t\bar t HH$
final state, we found a major number of events (Table~\ref{t5}) to
difference of the $b\bar b HH$ channel, but after considering the
decay products of both the Higgs boson (H) and the top quark $(t)$
to $b$ quarks, the expected final number of events will be very
small. Finally, the study of these processes is important and
could be useful to probe the triple Higgs boson self-couplings
$\lambda_{3H}$ given the following conditions: very high
luminosity, excellent $b$ tagging performance, and center-of-mass
energy in the range $\sqrt{s}=800-1000$ $GeV$, which is the most
favorable colliding energy for $b\bar b HH$ and $t\bar t HH$
production and for the lightest Higgs boson mass in the range
$M_H=110-130$ $GeV$. In addition, these results have never been
reported in the literature before and could be of relevance for
the scientific community.

\vspace{2.5cm}


\begin{center}
{\bf Acknowledgments}
\end{center}

We acknowledge support from CONACyT (Project: 78800, CB2008-01),
SNI and PROMEP (M\'exico). O. A. Sampayo would like to thank
CONICET (Argentina). We would also like to thank Maureen Sophia
Harkins for proofreading the manuscript.

\newpage

\end{document}